\documentclass[preprintnumbers,amsmath,amssymb,floatfix,12pt,prd,superscriptaddress,nofootinbib]{revtex4}
\usepackage{graphicx}
\usepackage{epsfig}
\usepackage{bm}
\usepackage{amsfonts}
\usepackage{color}
\begin{document}

\title{Accretion of Phantom Energy and Generalized Second Law of Thermodynamics for
Einstein-Maxwell-Gauss-Bonnet Black Hole}

\author{Mubasher Jamil}
\email{mjamil@camp.nust.edu.pk} \affiliation{Center for Advanced
Mathematics and Physics,\\ National University of Sciences and
Technology, H-12, Islamabad, Pakistan}

\author{Ibrar Hussain}
\email{ihussain@seecs.nust.edu.pk} \affiliation{School of Electrical Engineering and Computer Science,\\
National University of Sciences and Technology, H-12, Islamabad,
Pakistan}

\begin{abstract}

\vspace*{1.5cm} \centerline{\bf Abstract} \vspace*{.5cm} We have
investigated the accretion of phantom energy onto a 5-dimensional
\textcolor{black}{extreme} Einstein-Maxwell-Gauss-Bonnet (EMGB)
black hole. It is shown that the evolution of the EMGB black hole
mass due to phantom energy accretion depends only on the pressure
and density of the phantom energy and not on the black hole mass.
Further we study the generalized second law of thermodynamics (GSL)
at the event horizon and obtain a lower bound on the pressure of the
phantom energy.
\end{abstract}

 \maketitle
\newpage
\section{Introduction}
Various astronomical observations convincingly show \cite{perl} that
our universe is presently undergoing a phase of accelerated
expansion. Within the framework of General Relativity (GR), the
accelerated expansion of the universe could be explained by the
presence of a `cosmological constant' bearing negative pressure
which results in the stretching of the spacetime \cite{pad}. Many
other theoretical models have also been constructed to explain the
accelerated expansion of the universe including based on homogeneous
and time dependent scalar field like the quintessence
\cite{essence}, Chaplygin gas \cite{chaplygin} and phantom energy
\cite{phantom}, to name a few. The equation of state $p=\omega\rho$,
with $\omega<-1$, characterizes the phantom energy. It possesses
some weird properties: the cosmological parameters like scale factor
and energy density become infinite in a finite time; all
gravitationally bound objects lose mass with the accretion of
phantom energy; the fabric of spacetime is torn apart at the big
rip; and that it violates the standard relativistic energy
conditions. The astrophysical data coming from the microwave
background radiation categorically favors the phantom energy
\cite{cald}. Motivated from the dark energy models, we model phantom
energy by an ideal fluid with negative pressure.

Babichev et al \cite{babi} have shown that the mass of the black
hole will decrease with time when we consider the accretion of
phantom energy. \textcolor{black}{They showed that the mass will
vanish before the Big Rip.} After this seminal work the accretion of
dark energy onto a black hole have been investigated by many
authors. In GR, the accretion of phantom energy onto Schwarzschild
\cite{babi,jamil1}, Reissner-Nordstr\"{o}m (RN) \cite{jamil2},
Kerr-Newman (KN) \cite{jamil5} and primordial black holes has been
studied \cite{jamil3}. \textcolor{black}{In the case of RN black
hole, the mass of the black hole decreases but the electric charge
remains unaffected. Consequently the naked singularity appears at
the Big Rip. This henceforth violates the Penrose Cosmic Censorship
Hypothesis which forbids the existence of naked singularities. This
result also arises for the KN black hole. In another paper
\cite{milky}, the authors have investigated the accretion of phantom
energy on galaxies and deduce their destruction due to phantom
energy} The accretion of phantom energy on a 3-dimensional
\textcolor{black}{Banados-Teitelboim-Zanelli} (BTZ) black hole was
investigated in \cite{jamilakbar}. It was speculated there to
investigate the accretion dynamics in higher dimensional gravities
and modified theories of gravity.

The Einstein theory of gravity with Gauss-Bonnet (GB) term (given in
the next section) has some notable properties (see for example
\cite{cai,NDad}). The GB term attains nontrivial physical meaning in
5 dimensions \cite{NDad}. Consequently we investigate the accretion
of exotic phantom energy onto a static 5-dimensional EMGB black
hole. We show that the expression of the evolution of EMGB black
hole mass is independent of its mass and depends only on the energy
density and pressure of the phantom energy. It is well-known that
the horizon area of the black hole decreases with the accretion of
phantom energy \cite{GSL}, hence it is essential to study the GSL in
this case. We show that the validity of GSL in the present model
yields a lower bound on the phantom energy pressure. Beside, we
demonstrate that the first law of thermodynamics holds in the
present construction.

The plan of the paper is as follows: In second section we model the
accretion of phantom energy onto 5-dimensional EMGB black hole. In
third section, we study the GSL for EMGB black hole. Finally we
conclude our results.

\section{Model of accretion}

The action in 5-dimensional spacetime ($\mathfrak{M},g_{\mu\nu}$)
that represents the Einstein-Maxwell theory with a GB term and a
cosmological constant has the expression \cite{gauss,Bolu}
\begin{equation}\label{1}
S=\frac{1}{2}\int
d^5x\sqrt{-g}\Big[R-2\Lambda-\frac{1}{4}F_{\mu\nu}F^{\mu\nu}+\alpha
R_{GB}\Big],
\end{equation}
where
$R_{GB}=R^2-4R_{\mu\nu}R^{\mu\nu}+R_{\mu\nu\sigma\delta}R^{\mu\nu\sigma\delta}$,
is the GB term, while $\alpha$ is the GB coupling parameter having
dimension of $L^2$ ($\alpha^{-1}$ is related to the string tension
in Heterotic string theory), $\Lambda$ is a cosmological constant
and $F_{\mu\nu}$ is the electromagnetic field tensor. Variation of
the action (\ref{1}) with respect to the metric tensor yields the
EMGB field equations
\begin{equation}\label{1a}
G_{\mu\nu}-\alpha H_{\mu\nu}+\Lambda g_{\mu\nu}=T_{\mu\nu},
\end{equation}
where\textcolor{black}{
\begin{equation}
H_{\mu\nu}=2(RR_{\mu\nu}-2R_{\mu\lambda}R^\lambda_\nu-2R^{\gamma\delta}R_{\gamma\mu\delta\nu}
+R^{\sigma\gamma\delta}_\mu
R_{\sigma\nu\gamma\delta})-\frac{1}{2}g_{\mu\nu}R_{GB}.\nonumber
\end{equation}}
The spherically symmetric metric of a 5-dimensional EMGB black hole
is \cite{gauss,wils}
\begin{equation}\label{2}
ds^2=-f(r)dt^2+\frac{dr^2}{f(r)}+r^2d\Omega_3^2,
\end{equation}
where
\begin{equation}
d\Omega_3^2=(d\theta_1^2+\sin^2\theta_1(d\theta_2^2+
\sin^2\theta_2d\theta_3^2))\nonumber
\end{equation}
and
\begin{equation}\label{3}
f(r)=1+\frac{r^2}{4\alpha}-\frac{r^2}{4\alpha}\sqrt{1+\frac{16M\alpha}{\pi
r^4}-\frac{8Q^2\alpha}{3r^6}+\frac{4\Lambda\alpha}{3}},
\end{equation}
$M$ is the mass and $Q$ is the charge of the black hole. For extreme
EMGB black hole $M=Q$. The coefficient $g_{00}$ is termed as the
lapse function. The event horizon of the extreme EMGB black hole is
obtained by setting $f(r)=0$, and $df(r)/dr=0$ which turns out
\cite{gauss}
\begin{equation}\label{4}
r_e^2= \frac{1}{\Lambda}(1+2\cos\frac{\beta}{3}),
\end{equation}
where
\begin{equation}\label{5}
\cos\beta=1-\frac{Q^2\Lambda^2}{2},\quad Q^2\Lambda^2<4.
\end{equation}
Also we have$\sqrt{|g|}=r^3\sin^2\theta_1\sin\theta_2$, where $g$ is
the determinant of the metric. Here the apparent horizon is defined
by $A=4\Omega_4r^3$, where $\Omega_4=\pi^2/\Gamma(3)$ \cite{cai}.

To analyze the accretion of phantom energy onto the EMGB black hole,
we employ the formalism from the work by Babichev et al \cite{babi}.
The stress energy momentum tensor representing the phantom energy is
the perfect fluid
\begin{equation}\label{6}
T^{\mu\nu}=(\rho+p)u^\mu u^\nu+pg^{\mu\nu},
\end{equation}
where $\rho$ and $p$ are the energy density and pressure of the
phantom energy while $u^\mu=(u^0,u^1,0,0,0)$ is the velocity five
vector of the fluid flow. Also $u^1=u$ is the radial velocity of the
flow while the components $u^2$, $u^3$ and $u^4$ are zero due to
spherical symmetry of the EMGB black hole. Using the energy-momentum
conservation for $T^{\mu\nu}$ i.e. $T^{\mu\nu}_{;~\mu}=0$, we obtain
\begin{equation}\label{7}
u\sqrt{|g|}(\rho+p)\sqrt{f(r)+u^2}=C_1,
\end{equation}
where $C_1$ is constant of integration. Since the flow is inwards
the black hole therefore $u<0$. Also the projection of the energy
momentum conservation along the velocity five vector $u_\nu
T^{\mu\nu}_{;\mu}=0$ (the energy flux equation) is
\begin{equation}\label{8}
u\sqrt{|g|}\exp\Big[\int\limits_{\rho_\infty}^{\rho_h}\frac{d\rho}{\rho+p}\Big]=-A_1.
\end{equation}
Here $A_1$ is an integration constant and the associated minus sign
is taken for convenience. Also $\rho_h$ and $\rho_\infty$ are the
energy densities of phantom energy at the EMGB horizon and at
infinity respectively. From (\ref{7}) and (\ref{8}), we obtain
\begin{equation}\label{9}
(\rho+p)\sqrt{f(r)+u^2}\exp\Big[-\int\limits_{\rho_\infty}^{\rho_h}\frac{d\rho}{\rho+p}\Big]=C_2,
\end{equation}
where $C_2=-C_1/A_1=\rho_\infty+p(\rho_\infty)$. The rate of change
in the mass of black hole $\dot M=-2\pi^2 r^3 T^1_0$, is given by
\begin{equation}\label{10}
dM=2\pi^2 A_1(\rho_\infty+p_\infty)dt.
\end{equation}
Note that $\rho_\infty+p_\infty<0$ (violation of null energy
condition) leads to decrease in the mass of the black hole.
Moreover, the above expression is also independent of mass contrary
to the Schwarzschild, Reissner-Nordstr\"{o}m and Kerr-Newman black
holes \cite{jamil1,jamil2,jamil5}. Further, the last equation is
valid for any general $\rho$ and $p$ violating the null energy
condition, thus we write
\begin{equation}\label{11}
dM=2\pi^2 A_1(\rho+p)dt.
\end{equation}

\section{ Generalized second law of thermodynamics and EMGB black hole}

In this section we discuss the thermodynamic of phantom energy
accretion that crosses the event horizon of the EMGB black hole. Let
us first write the EMGB metric in the form
\begin{equation}\label{12}
ds^{2} = h_{mn}dx^{m}dx^{n}+r^{2}d\Omega_3^{2}, \ \ \ m,n=0,1
\end{equation}
where $h_{mn}= \text{diag}(-f(r), 1/f(r))$, is a 2-dimensional
metric. From the condition of normalized velocities $u^{\mu}u_{\mu}
= -1$, we obtain the relations
\begin{equation}\label{13}
u^{0} = f(r)^{-1}\sqrt{f(r) + u^{2}}, ~~~ u_{0} = -\sqrt{f(r) +
u^{2}}.
\end{equation}
The components of stress energy tensor are
\begin{equation}\label{14}
T^{00}=f(r)^{-1}[(\rho + p)(\frac{f(r)+u^{2}}{f(r)})-p],
\end{equation}
and
\begin{equation}\label{15}
T^{11}=(\rho + p)u^2+f(r)p.
\end{equation}
With the help of (15) and (16) we calculate the work density which
is defined by $W=-\frac{1}{2}T^{mn}h_{mn}$ \cite{cai}. It comes out
\begin{equation}\label{16}
W=\frac{1}{2}(\rho - p).
\end{equation}
The energy supply vector is defined by
\begin{equation}\label{17}
\Psi_{n}=T^{m}_{n}\partial_{m}r + W\partial_{n}r.
\end{equation}
The components of the energy supply vector are
\begin{equation}\label{18}
\Psi_{0}= T^{1}_{0}=-u(\rho+p)\sqrt{f(r)+u^2},
\end{equation}
and
\begin{equation}\label{19}
\Psi_{1}=T^{1}_{1}+W=(\rho+p)\Big(\frac{1}{2}+\frac{u^2}{f(r)}\Big).
\end{equation}
The change of energy across the apparent horizon is determined
through $-dE\equiv-A\Psi$, where $\Psi=\Psi_0dt+\Psi_1dr$. The
energy crossing the event horizon of the EMGB black hole is given by
\begin{equation}\label{20}
dE=2\pi^2 r_e^3u^2(\rho+p)dt.
\end{equation}
Assuming $E=M$ and comparing (\ref{11}) and (\ref{20}), we can
determine the value of constant $A_1=u^2\sqrt{|g|}$.

The entropy of the EMGB black hole is \cite{cai}
\begin{equation}\label{21}
S_h=\frac{\pi^2r_e^3}{2}(1+\frac{12\alpha}{r_e^2}) .
\end{equation}
It can be shown easily that the thermal quantities, change of
phantom energy $dE$, horizon entropy $S_h$ and horizon temperature
$T_h$ satisfy the first law $dE=T_hdS_h$, of thermodynamics. After
differentiation of last equation w.r.t. $t$, and using (\ref{11}),
we have
\begin{equation}\label{22}
\dot{S}_h=\pi u^2(\rho+p)r_e^3.
\end{equation}
Since all the parameters are positive in (\ref{22}) except that
$\rho+p<0$, it shows that the second law of thermodynamics is
violated i.e. $\dot{S}_h<0$, as a result of accretion of phantom
energy on the EMGB black hole.\\ Now we proceed to the GSL. It is
defined by \cite{shey}
\begin{equation}\label{23}
\dot{S}_{tot}=\dot{S}_h+\dot{S}_{ph}\geq0.
\end{equation}
In other words, the sum of the rate of change of entropies of black
hole horizon and phantom energy must be positive. We consider event
horizon of the EMGB black hole as a boundary of thermal system and
the total matter energy within the event horizon is the mass of the
EMGB black hole. We also assume that the horizon temperature is in
equilibrium with the temperature of the matter-energy enclosed by
the event horizon, i.e. $T_h=T_{ph}=T$, where $T_{ph}$ is the
temperature of the phantom energy. Similar assumptions for the
temperatures $T_h$ and  $T_{ph}$ has been studied in \cite{davies}.
We know that the Einstein field equations satisfy first law of
thermodynamics $T_hdS_h=pdV+dE$, at the event horizon \cite{azad}.
We also assume that the matter-energy enclosed by the event horizon
of the EMGB black hole also satisfy the first law of thermodynamics
given by
\begin{equation}\label{24}
T_{ph}dS_{ph}=pdV+dE.
\end{equation}
Here the horizon temperature is given by \cite{cai}
\begin{equation}\label{25}
T_h=\frac{1}{2\pi r_e}.
\end{equation}
In this paper, we are assuming that $T_h=T_{ph}=T$. Therefore
(\ref{23}) gives
\begin{equation}\label{26}
T\dot{S}_{tot}=T(\dot{S}_h+\dot{S}_{ph})=4\pi^2
u^2(\rho+p)r_e^3(1-p\pi^2 r_e^2k(r_e)),
\end{equation}
where
\begin{equation}\label{27}
k(r_e)=-\frac{\sqrt{{1-(\frac{r_e^2
\Lambda-1}{2})^2}}}{3\sqrt{4-\Lambda^2 Q^2}}.
\end{equation}
From the above equation, note that $u^2>0$, $r_e^3>0$ and
$\rho+p<0$. The GSL holds provided $1-p\pi^2 r_e^3k(r)<0$ which
implies
\begin{equation}\label{28}
p>\frac{1}{\pi^2 r_e^3 k(r_e)}.
\end{equation}
Since the pressure of the phantom energy is negative ($p<0$),
therefore the GSL gives us the lower bound on the pressure of the
phantom energy.
\begin{equation}\label{29}
-\frac{3\sqrt{4-\Lambda^2Q^2}}{\pi^2
r_e^2\sqrt{1-(\frac{r_e^2\Lambda-1}{2})^2}}< p<0.
\end{equation}
The GSL in the phantom energy accretion holds within the inequality
(\ref{29}) which is independent of the GB parameter. Otherwise the
GSL does not hold which forbid evaporation of EMGB black hole by the
phantom accretion \cite{pavon}. In addition, it is not clear whether
the GSL should be valid in presence of the phantom fluid not
respecting the dominant energy condition \cite{pavon}.

\section{Conclusion}

In this paper, we have studied the accretion of exotic phantom
energy onto extreme EMGB black hole. The motivation behind this work
is to investigate the accretion dynamics in higher dimensional
gravity. Our analysis has shown that evolution of mass of the EMGB
black hole would be independent of its mass and will be dependent
only on the energy density and pressure of the phantom energy in its
vicinity. Due to spherical symmetry, the accretion process is simple
since the phantom energy falls radially on the black hole. This
result is similar to the one obtained for the BTZ black hole
\cite{jamilakbar}. \textcolor{black}{Since the result of the
accretion of phantom energy in 4-dimensions (for the Schwarzschild,
RN and KN black holes) is mass dependent and in the case of
3-dimension BTZ and 5-dimensional EMGB is mass independent. This
raises the question whether this dependency of $\dot M$ on mass is
restricted to black holes in 4-dimensions only. Therefore we
conjecture the following: \textit{the rate of change of mass of a
black hole due to the phantom energy depends on the mass of a black
hole in 4-dimensions only.}}

\textcolor{black}{ We also studied GSL for the EMGB black hole. This
is performed on the assumption that the event horizon of EMGB black
hole acts as a boundary of the thermal system. Moreover the phantom
energy crosses the event horizon radially and reduces the mass of
the black hole. Furthermore the black hole horizon is in thermal
equilibrium with the phantom energy falling on the black hole. Under
these assumptions we deduced that the GSL holds provided the
pressure of the phantom energy $p$ acts as the lower bound
(\ref{29}) on the black hole parameters.}

An interesting question is that what would be the fate of the black
hole near the big rip. For a Schwarzschild black hole \cite{babi},
the mass completely disappears without any remnant, while for the
Reissner-Nordstr\"{o}m black hole \cite{jamil2}, a remnant in the
form of naked singularity appears because the accretion does not
effect the charge. In the present scenario of an extremal EMGB black
hole, \textcolor{black}{ for which mass and charge are on equal
footing,} the accretion leaves {\it no} remnant analogous to a
Schwarzschild black hole. \textcolor{black}{ We emphasize that the
present study cannot be reduced to that for Schwarzschild solution
by choosing $\alpha=0$ since the EMGB solution (\ref{3}) becomes
undefined.}

\subsubsection*{Acknowledgment}
\textcolor{black}{We would like to thank the referees for giving
useful comment to improve this paper. }

\end{document}